\documentclass[12pt]{article}
\usepackage{latexsym,graphicx}
\newcommand{\be}{\begin{equation}}
\newcommand{\ee}{\end{equation}}
\usepackage[left=2.5cm,top=2.5cm,right=2.5cm,bottom=1.5cm]{geometry}
\usepackage{float}
\usepackage{epstopdf}
\usepackage{cite}
\begin{document}
\begin{center}
\large{\bf{Traversable wormhole solutions with non-exotic fluid in  framework of $f(Q)$ gravity}}\\
\vspace{10mm}
\normalsize{Umesh Kumar Sharma$^1$, Shweta$^2$, Ambuj Kumar Mishra$^3$}  \\
\vspace{5mm}
\normalsize{$^{1,2,3}$Department of Mathematics, Institute of Applied Sciences and Humanities, GLA University,
	Mathura-281 406, Uttar Pradesh, India}\\

\vspace{2mm}
$^1$E-mail: sharma.umesh@gla.ac.in\\
\vspace{2mm}
$^2$E-mail: shwetaibs84@gmail.com  \\
 \vspace{2mm}
$^3$E-mail: ambuj\_math@rediffmail.com \\
\vspace{2mm}
\end{center}
\vspace{10mm}
\begin{abstract}
	The presence of exotic matter for the existence of the wormhole geometry has been an unavoidable problem in GR. In recent studies researchers have tried to deal with this issue using  modified gravity theories where the WH geometry is explained by the extra curvature terms and NEC's are not violated signifying the standard matter in the WH geometry. In the present article we are trying to find the solutions of traversable wormholes with normal matter in the throat within the framework of symmetric teleparallel gravity $f(Q)$ where $Q$ is the non metricity scalar which defines the gravitational interaction. We will examine the wormhole geometries for three forms of function $f(Q)$. First is the linear form $f(Q)=\alpha Q$, second a non -linear form $f(Q)=\alpha Q^2 + \beta$ and third one a more general quadratic form $f(Q)=\alpha Q^2 + \beta Q + \gamma$ with $\alpha$, $\beta$ and $\gamma$ being the constants. For all the three cases the shape function is taken as  $b(r) = {\frac{r_{0}\ln(r+1)}{\ln({r_0}+1)}}$ where $r_0$ is the throat radius. A special variable redshift function is considered for the discussion. All the energy conditions are then examined for the existence and the stability of the wormhole geometry. 
	
\end{abstract}



\section{Introduction}

The Universe and its evolution has been a great mystery for mankind and in the recent past tremendous work has been done and is going on to unveil the secrets of the early Universe. When we take a look at the stars in the sky, we are actually looking into the past. Many techniques such as EHT, VIRGO, LIGO and many more are used to gather the observational data which is used to explore the gravitational properties in strong fields \cite{ref1,ref2,ref3,ref4}. Ever since humans are amused if they could travel to the  far existing galaxies or if there exists any life far beyond in space-time. For such fantasies there came the concept of traversable wormholes which gained a special attention among the researchers. The idea of wormholes i.e. the supposed connection of the shape of a tube or tunnel between two asymptotically flat regions in space-time was first conceived by Flamm as Schwarschild solutions \cite {ref5}. These are cosmological entities without any singularity or event horizon \cite{ref6} which leads to the study of non-standard matter and the contribution of gravity in its formation. In 1935 Einstein and Rosen gave the concept of Einstein-Rosen bridge that was the extension of the static wormhole \cite{ref7}. In GR and modified gravity theories the solution of field equations gives the geometry of the wormhole as a shortcut between distant Universes \cite{ref8}. Eventually the wormhole solutions imitating the Einstein-Rosen bridge were obtained by connecting two Schwarschild solutions \cite{ref9,ref10}. Later on, a spherically symmetric traversable wormhole was given by Morrice and Thorne \cite {ref11} which were subsequently seen to be in tune with tachyonic massless scalar field \cite{ref12,ref13}. Afterwards, with the help of Yurtsever, Morrice and Thorne put forward the idea of a time machine by exploring the plausible solution of traversable wormholes through which the matter and radiation can pass \cite{ref14,ref15,ref16}.\\

The focus nowadays is on exploring the traversable wormholes without any singularity or horizon . To treat this issue, a presumption can be made on the line element. In addition to this, the conditions can be imposed on the WH throat using Birkhoff theorem \cite{ref11,ref17} so that the radial tension exceeds the mass-energy density. Consequently the NEC's are spoiled at the throat by the energy -momentum tensor \cite{ref18,ref19,ref20,ref21}.
This leads to the existence of the phantom fluid or phantom energy which signifies the accelerated expansion of the Universe. To mitigate this problem several theories have been given. But in this regard, either the non-standard fluid can be considered or the modified gravity theories can be taken to explain  the  WH geometry by existence of higher order curvature terms. The traversable wormholes and thin shell wormholes are explored in this regard in $f(R)$ gravity. With the coupling of geometry with matter terms and obtaining various relations of  radial and tangential pressure, the WHs are examined in $f(R,T)$ gravity \cite{ref21a,ref22} where  energy conditions are examined against different shape functions. Many researchers inspected the wormhole solutions in $f(R,T)$ and $R^2$ gravity \cite{ref22a,ref22b,ref22c,ref22d,ref22e,ref22f,ref22g}. The wormhole geometries are examined in teleparallel and other extended gravitational theories \cite{ref23,ref23a,ref23b,ref23c,ref23d}. The teleparallel gravity is considered to be an alternative to GR where Torsion $T$ defines the gravitational interaction \cite{ref24}.  Hence, there are several inspirations to investigate speculations beyond the standard definition of gravity.\\

In our present work,  we  examine the wormhole solutions in backdrop of symmetric teleparallel gravity $f(Q)$ proposed by Jimenez et al. \cite{ref27},  where the geometry is torsion and curvature free and the non metricity term $Q$ only defines the gravitational interactions.  An observational analysis of  different forms of $f(Q)$ gravity for the validity of these models has been performed considering several observational probes like Cosmic Microwave Background distance priors, Gamma
	Ray Bursts, Quasars,  Type Ia Supernovae, the expansion rate data from early-type galaxies and Baryon Acoustic Oscillations data, where the Lagrangian of  $f(Q)$ gravity  is reformulated as an explicit function of the redshift,
	$f(z)$  \cite{ref28e,ref28f}. A novel model has been proposed in the framework of $f(Q)$ gravity, which is a class of  gravitational modification
	emerging from the assimilation of non-metricity  \cite{ref28g}. Mandal et al. \cite{ref28h,ref28i} presented a complete
	test of energy conditions and 
	cosmographical approach  for $f(Q)$  gravity models. Harko et al. \cite{ref28j}  presented an
	expansion  of  $f(Q)$  gravity, by presenting a new class of theories where $Q$
	is coupled non-minimally to the matter Lagrangian, in
	the context of the metric-affine formalism. Hassan et al. \cite{ref28,ref28a1} have explored the wormhole geometries in the framework of $f(Q)$ gravity. They discussed three types of wormhole geometries for two different function forms, linear and non-linear of $f(Q)$. They obtained wormhole geometries with exotic matter in the throat which instigated us and gave motivation to find a wormhole solution with normal matter. Taking into account that in GR the gravitational and inertial reactions can not be differentiated but by rearranging the frames in teleparallel theory the gravitational theory can be described co-variantly \cite{ref25,ref26}, and the authors introduced $f(Q)$ symmetric teleparallel gravity theory by extending GR as a newer GR \cite{ref27}. Since the presence of exotic matter is supposed to be the unrealistic approach, this problem is the motivation of our work. \\

The formation of this manuscript is as follows:
In section (2) we have formulated the $f(Q)$ gravity. The basic constraints for the shape function and conditions required for a traversable wormhole are depicted in section (3). The field equations under $f(Q)$ gravity are also described in this section. We have discussed the energy conditions and their implications in subsection (3.1). In section (4) and its subsections,  the wormhole solutions are obtained under three functional forms of $f(Q)$. The discussion and final remarks are given in section (5).

\section{The $f(Q)$ Gravity} 

The action for symmetric teleparallel gravity as suggested by Jimenez et al. \cite{ref27, ref28a} is given as

\begin{equation}\label{eq1}
	S = \int \left[\frac{1}{2} f\left(Q \right) + {\mathcal{L}_{\mu}} \right] \sqrt{-g}  d^4 x  ,  
\end{equation}

where, the function of the non-metricity term Q is taken as $f(Q)$ and the Lagrangian density of matter is given by $ \mathcal{L}_{\mu}$. Also the determinant of the metric ${g}_{\eta \zeta}$ is given by $g$.
The non-metricity tensor is given by

\begin{equation}\label{eq2}
	{Q}_{\lambda \eta \zeta} = \nabla _{\lambda \eta \zeta},  
\end{equation}

There are two independent traces of non-metricity tensor, which we signify as

\begin{equation}\label{eq3}
	{Q}_\phi = {Q_\phi} ^{\eta} _{\eta}, \qquad\qquad  \bar{Q}_\phi = Q^{\eta}_{\phi \eta}.
\end{equation}

The non-metricity conjugate, analogous to super potential of so called New GR \cite{ref27} can be written as,

\begin{equation}\label{eq4}
	{P^\phi}_{\eta \zeta} = \frac{1}{4} \left[ - {Q^\phi}_{\eta \zeta} + 2 {Q ^ \phi}_{\eta _\zeta}  + {Q ^ \phi} {g}_{\eta \zeta} - \bar{Q}^{\phi} {g}_{\eta \zeta}  - {\delta^\phi}_{\left(\eta  Q_\zeta \right)} \right],
\end{equation}
which is obtained by taking the trace of non-metricity tensor of the form

\begin{equation}\label{eq5}
	Q =  - Q _ {\phi \eta \zeta} P ^{\phi \eta \zeta}.
\end{equation}

Now, the energy-momentum tensor, explaining the nature of matter filled in the space-time is defined as 

\begin{equation}\label{eq6}
	T _ {\eta \zeta} = - \frac{2}{\sqrt{-g}} \frac{ \delta \left(\sqrt{-g} \mathcal{L}_{\mu} \right)}{ \delta g^{\eta \zeta}}.
\end{equation} 

On varying the action (\ref{eq1}) with respect to the metric tensor $g _ {\eta \zeta}$, we get the motion equations given by  

\begin{equation}\label{eq7}
	\frac{2 \nabla _ \chi}{\sqrt{- g }} \left( \sqrt {-g} {f_Q} {P^\chi} _{\eta \zeta} \right) + \frac{1}{2} g_{\eta \zeta} f + {f_Q} \left( P _{\eta \chi \psi} {{Q_\zeta} ^\chi \psi} - 2 {Q}_{\chi \psi \eta}  {{P^\chi \psi} _ \zeta} \right) = - {T} _{\eta \zeta},
\end{equation} 

where $ {f}_{Q} $ is the total derivative of $f$ with respect to $Q$. One can also obtain following equation, while varying (\ref{eq1}) with respect to the connections,
\begin{equation}\label{eq8}
	{\nabla _ \eta}{\nabla _ \zeta} \left( \sqrt{-g} {f_Q} {P^\chi} _{\eta \zeta} \right) = 0.
\end{equation}

\section{Wormhole Geometry and Solution of Field Equations in Symmetric Teleparallel Gravity i. e. $f(Q)$ Gravity}

To study the wormhole geometry in $f(Q)$ gravity, here we  take the Morris Thorne class general spherically symmetric static geometry given by the line element

\begin{equation}\label{eq9}
	ds^2= -\exp (2 \phi(r)) dt^2 + \left (\frac{r - b \left( r \right)}{ r} \right) ^{-1} d r^2 +{r^2} d {\theta}^2 + {r^2} {\sin}^2 {\theta} d{\phi}^2.
\end{equation}
Here $\phi (r)$ is said to be the redshift function of the intrusive object in the radial coordinate $r$ where $ 0 < r_0 \leq r \leq \infty$. To prevent from the event horizons or the appearance of any singularity about the throat, the value of redshift function must be finite everywhere and should not vanish at the throat. In our present work we will discuss the wormhole geometry by taking a variable redshift function. The function $b(r)$, prominently known as shape function, ascertains the wormhole shape. Therefore, the shape function $b(r)$ must abide by some constraints to stand with the wormhole geometry. The conditions $0 < 1-\frac{b(r)}{r}$ and $b(r_0)= r_0$ where $r_0 \leq r \leq \infty$  must be satisfied by the shape function.  These conditions are known as throat conditions. Here $r_0$ is the minimum value of $r$ and is the throat radius. The flaring out condition, which signifies the minimum size of the throat is required to maintain the geometry of the wormhole. The flaring out condition can be given as $b'(r_0) <1$. The flaring out condition establishes the traversability through wormhole space time. In addition with above conditions another vital condition for the appropriate shape function is the asymptotically flatness condition i. e. $\frac{b(r)}{r}\rightarrow 0  \qquad as \qquad |r|\rightarrow \infty$ which the shape function must obey.
One more crucial criterion for the traversable wormhole is that the proper radial distance given by $x(r)= \pm \int_{r_0}^{r} \frac{dr}{\sqrt{\frac{r - b(r)}{r}}}$ has to be finite corresponding to the radial coordinate. Therefore, it is a decreasing function falling from the upper Universe at  $x = + \infty$ to the throat of the wormhole at $x=0$ and then further going down to $x=- \infty$.
Also, the proper radial distance $x(r)$ can not be less the radial coordinate distance i. e. $ r - r_0 \leq |x(r)|$. The wormhole throat connects the upper and lower regions of throat represented by the positive and negative values of $x$ respectively.\\

Here, we consider that the wormhole throat is filled with an anisotropic fluid that accounts for the following stress-energy-momentum tensor

\begin{equation}\label{eq10}
	T_\eta ^\zeta = \left(\rho + p_t\right) {u_\eta} {u^\zeta} - {p_t}{\delta_\eta ^\zeta} + \left(p_r - p_t \right){v_\eta} {v^\zeta}.
\end{equation} 

Here, $v_\eta$ represents the unitary space-like vector which is in radial direction and four velocity is given by $u_\eta$. $\rho$, $p_r$ and $p_t$ are the energy density, radial pressure and the tangential pressure respectively.
Within the framework of $f(Q)$  gravity, for the line element (\ref{eq9}), the non-metricity tensor has the trace Q of the form

\begin{equation}\label{eq11}
	Q = - {\frac{2}{r^3}} \left[ r - b(r) \right] \left[ 2.r.{\phi'(r)} + 1\right].
\end{equation} 

Now solving (\ref{eq7}) with (\ref{eq9}) and (\ref{eq10}), we can get the following values of energy density,  radial pressure and tangential pressure in terms of $r$.

\begin{equation}\label{eq12}
	\rho =	\left[ {\frac{1}{r^3}} \left( {r} - r b'(r) - b(r) + 2 r \phi '(r) \left(r - b(r) \right) \right) \right] f_Q + \frac{2}{r^2}\left(r - b(r) \right) f_Q + \frac{f}{2}, 
\end{equation} 

\begin{equation}\label{eq13}
	p_r = - \left[ {\frac{2}{r^3}} \left(r - b(r)\right) \left( 2 r \phi'(r) +1\right) -1 \right] f_Q -\frac{f}{2} ,
\end{equation} 

\begin{eqnarray}\label{eq14}
	p_t &=& - \left[ {\frac{1}{r^3}}\left( \left[r - b(r)\right] \left[ 1 +  r \phi'(r)\left(3 + r \phi'(r)\right) +r^2 \phi''(r)\right] - \frac{1}{2}\left[r b'(r) - b(r) \right]\left[ 1 + r \phi'(r)\right] \right) \right]f_Q \nonumber\\
	&-&
	\frac{1}{r^2}\left[r - b(r)\right]\left[1 + r \phi'(r)\right]f_Q - \frac{f}{2}.
\end{eqnarray} 

By choosing an appropriate value of shape function as $b(r)$, one can investigate the wormhole geometry.

\subsection{The Energy conditions}
The energy conditions are some necessary constraints that are derived from the Raichaudhuri equations which describe the temporal evolution for the time-like vector $u^\eta$ and the null geodesics $k_\eta$ as \cite{ref29}

\begin{equation}\label{15}
	\frac{d\theta}{d\tau} - \omega _{\eta\zeta}\omega^{\eta\zeta} + \sigma_{\eta\zeta}\sigma^{\eta\zeta} + \frac{1}{3} \theta^2 + R_{\eta\zeta} u^\eta u_\zeta = 0,
\end{equation}

\begin{equation}\label{16}
	\frac{d\theta}{d\tau} - \omega _{\eta\zeta}\omega^{\eta\zeta} + \sigma_{\eta\zeta}\sigma^{\eta\zeta} + \frac{1}{2} \theta^2 + R_{\eta\zeta} k^\eta k_\zeta = 0.
\end{equation}

Here $k^\eta$ represents the vector field while the shear or spatial tensor is expressed as $R_{\eta\zeta} k^\eta k_\zeta$ having  $\sigma^{2}= \sigma_{\eta\zeta}\sigma^{\eta\zeta} \geq 0$ and $\omega_{\eta\zeta}\equiv0$.
These constraints or  energy conditions, in terms of energy density,  radial pressure and tangential pressure are accountable for the existence and stability of the traversable wormholes. These conditions give an insight into the nature of matter that is filled in the throat. The Rai Chaudhary conditions comply with the following conditions in case of attractive geometry i.e. for $\theta<0$

\begin{equation}\label{17}
	R_{\eta\zeta} u^\eta u_\zeta \geq 0,
\end{equation}

\begin{equation}\label{18} 
	R_{\eta\zeta} k^\eta k_\zeta \geq 0.
\end{equation}

For the anisotropic fluid, the energy conditions can be given as
 
		
 \begin{itemize}  	
	\item  Null energy condition or NEC: This condition implies the non-negativity of principle pressures i.e. $\forall i, \rho(r) + p_{i}\geq 0 \Leftrightarrow$ NEC. In tensor form NEC is given as $T_{\eta\zeta} k^{\eta}k^{\zeta} \geq 0$.   	
	\item
	Weak energy condition or WEC: $\rho(r)\geq 0$ and $\forall i, \rho(r) + p_{i}\geq 0$ or $T_{\eta\zeta} k^{\eta}k^{\zeta} \geq 0$ for a time like vector describes the positivity of energy density locally.  	
	\item
	Strong energy condition or SEC: $ (T_{\eta\zeta}-\frac{T}{2}g_{\eta\zeta}) k^{\eta}k^{\zeta} \geq 0 $ or in terms of principle pressures $ T = -\rho(r) + \sum_{j} p_{j}$ and  $\forall j, \rho(r) + p_{j}\geq 0,  \rho(r) +\sum_{j} p_{j}\geq 0 $.
	The violation of strong energy condition is a must to acknowledge the Universe inflation.  	
	\item
	Dominant energy condition or DEC: $ T_{\eta\zeta} k^{\eta}k^{\zeta} \geq 0$ where $T_{\eta\zeta}k^{\eta}$ is not space-like or in terms of $p_r$ and $p_t$, DEC is given as  $\rho(r) \geq 0$ and $\forall i, \rho \pm p_{i} \geq0$. The DEC shows that the speed of light is the maximum,  which energy transfer can achieve and flow of energy or mass can not exceed the speed of light.
\end{itemize}

\section{Solution of the Wormhole Models} 

In our present work, we  explore the wormhole solution for three different $f(Q)$ forms. First we will examine the shape function for the linear function of $Q$, second we will take the non-linear form of $f(Q)$ and in third part we will investigate the wormhole geometry for the general quadratic function of $Q$.
For each case we are investigating the logarithmic shape function
$b(r)=\frac{r_0 \ln (r+1)}{\ln (r_0 +1)}$ with  $r_0$ being the throat radius \cite{ref28b, ref28c}. As we have already mentioned in the previous section, the redshift function must attain non-zero finite value to avoid any singularity and horizons around the throat. Here in our work, we take variable redshift function $\phi(r)=\ln \left(\frac{r_0}{r}+1\right)$ to validate the asymptotically flatness of wormhole geometry \cite{ref28d}. The equation of state parameter which gives the nature of the matter fluid filled in the throat, can be given in the terms of radial pressure $p_r$ and energy density $p_t$ as 

\begin{equation}\label{19} 
	p_r = \omega\rho.
\end{equation}

The EoS plays a crucial part in defining the wormhole geometry by giving a clue about the throat matter. For $ -1<\omega<-1/3$, the associated matter corresponds to the quintessence of dark energy which signifies the acceleration of the expansion of the Universe. $\omega < -1$ for phantom fluid is considered a leading candidate for dark energy which also accelerates the Universe expansion.

\begin{figure}
	(a)\includegraphics[width=8cm, height=8cm, angle=0]{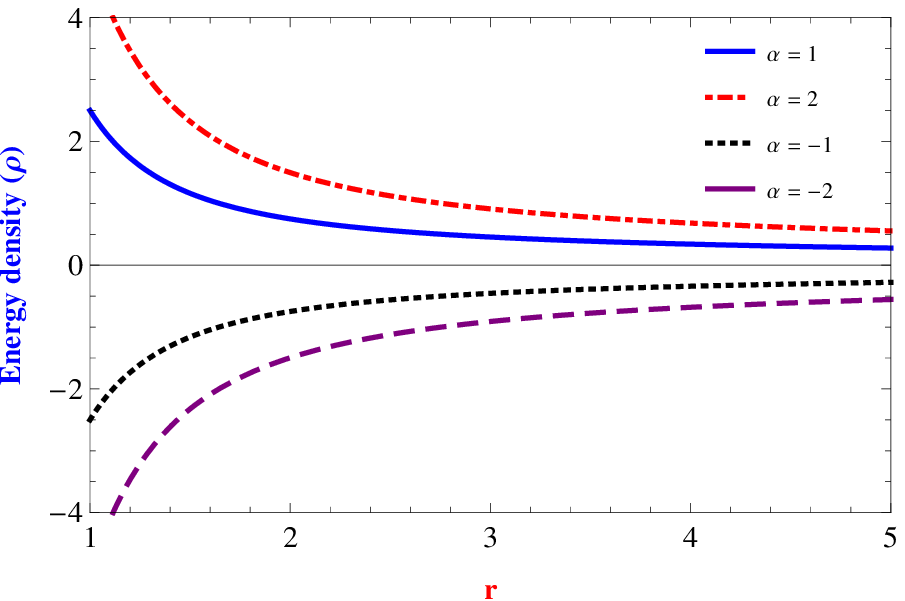}
	(b)\includegraphics[width=8cm, height=8cm, angle=0]{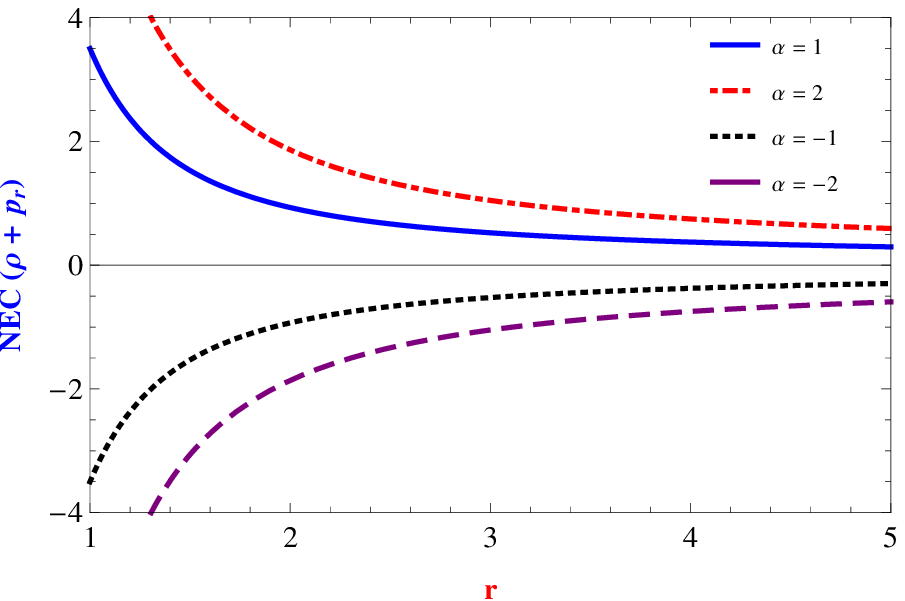}
	\caption {Variation of Energy density ($\rho$) and NEC ($\rho + p_r$) for throat radius $r_0 = 1$ }
\end{figure}

\begin{figure}
	(a)\includegraphics[width=8cm, height=8cm, angle=0]{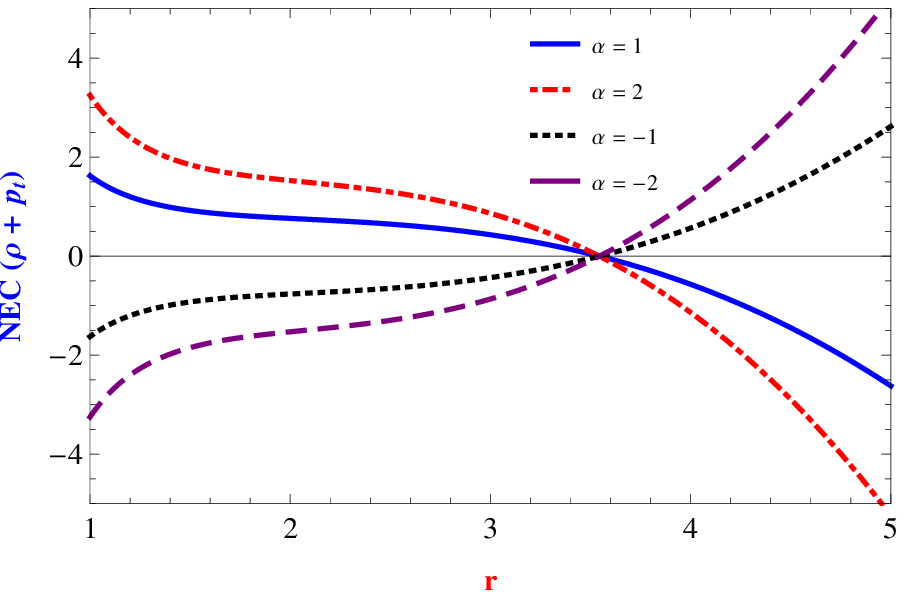}
	(b)\includegraphics[width=8cm, height=8cm, angle=0]{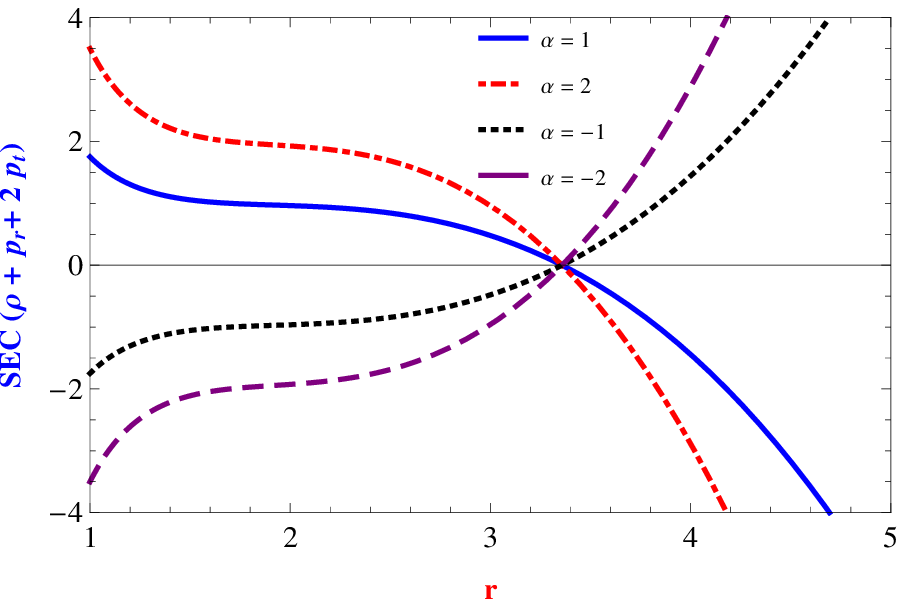}
	\caption {Variation of NEC ($\rho+p_t$) and  SEC ($\rho + p_r+ 2p_t$) for throat radius $r_0 = 1$.}
\end{figure}

\subsection{Wormhole solutions for linear form of $f(Q)$ i.e. $f(Q)=\alpha Q$}
In this section, we have taken the linear form $f(Q)= \alpha Q$. Here the constant $\alpha$ denotes the teleparallel gravitational term. This form of fetches the symmetric teleparallel equivalent of GR and helps in comparing the wormhole solutions to the fundamental ones. Inducing the above logarithmic shape function $b(r)$ and considering specific variable redshift function $\phi(r)$ within equations  (\ref{eq12}),(\ref{eq13}) and (\ref{eq14}), we obtain the energy density $\rho$, radial pressure $p_r$ and tangential pressure $p_t$  as  

\begin{equation}\label{eq20}
	\rho = {\frac { \left( r \left( r+1 \right)  \left( {r}^{2}+r_{{0}}r+2\,r_{{0}} \right) \ln  \left( r_{{0}}+1 \right) - \left(  \left( r+1 \right)  \left( {r}^{2}+r_{{0}}r+2\,r_{{0}} \right) \ln  \left( r+1 \right) +1/2\,r \left( r_{{0}}+r \right)  \right) r_{{0}} \right) \alpha}{2{r}^{3}\ln  \left( r_{{0}}+1 \right)  \left( r_{{0}}+r
			\right)  \left( r+1 \right) }}
\end{equation}  

\begin{equation}\label{eq21}
	p_r = {\frac {\alpha\, \left( 2\,\ln  \left( r_{{0}}+1 \right) r+\ln 
			\left( r+1 \right)  \left( -r_{{0}}+r \right)  \right) r_{{0}}}{{r}^{
				3}\ln  \left( r_{{0}}+1 \right)  \left( r_{{0}}+r \right) }}
\end{equation}

\begin{equation}\label{eq22}
	p_t= -{\frac {\alpha\, \left( r \left( r+1 \right)  \left( {r}^{2}+r_{{0}}
			\right) \ln  \left( r_{{0}}+1 \right) - \left(  \left( r+1 \right) 
			\left( {r}^{2}-r+r_{{0}} \right) \ln  \left( r+1 \right) +{r}^{2}
			\right) r_{{0}} \right) }{{r}^{3}\ln  \left( r_{{0}}+1 \right) 
			\left( r_{{0}}+r \right)  \left( r+1 \right) }}
\end{equation}

To analyze the wormhole geometry for this case, the four energy conditions, as discussed in above, have been plotted in Fig. 1 - 4. The  energy density $\rho$ is plotted against the radial coordinate $r$ and different values of constant $\alpha$ in the Fig. 1(a). As we can see from the figure, energy density is positive throughout the region for positive values of $\alpha$ i. e. for $\alpha = 1, 2$ and negative for the negative values of $\alpha$ i. e. $\alpha = -1,-2$. We see that radial NEC as depicted in Fig. 1(b), is satisfied for positive values of $\alpha$ and  for every values of $r\geq r_0$  and for negative $\alpha$ the $\rho + p_r$ is found negative for everywhere. On the other hand, for positive values of $\alpha$, the tangential NEC in Fig. 2(a) is satisfied at the throat and everywhere for $r_0 \leq r \leq 3.54$  and is violated as $r$ increases from 3.54. Also for negative $\alpha$ tangential NEC is violated for $r_0\leq r \leq 3.54$ and satisfied for $r \geq 3.54$. Hence we can say that null energy conditions are satisfied for positive values of $\alpha$ and for $r_0 \leq r\leq 3.54$. The SEC can be observed from Fig. 2(b) which is obeyed for positive $\alpha$ and $r_0 \leq r \leq 3.37$ and for negative $\alpha$, SEC is violated for this region. One can observe from Fig. 3 that both the DEC's are also satisfied at the throat for positive $\alpha$. As can be seen from Fig. 4(b), The anisotropy parameter is negative throughout the space time for positive values of $\alpha$  while for negative $\alpha$ anisotropy parameter is positive for all $r$.
The equation of state parameter gives the nature of fluid in space-time. From Fig. 4(a), it is observed that the value of the EoS parameter $\omega$ is positive hence $\omega >-\frac{1}{3}$ which indicates the presence of only ordinary matter at the throat.

\begin{figure}
	(a)\includegraphics[width=8cm, height=8cm, angle=0]{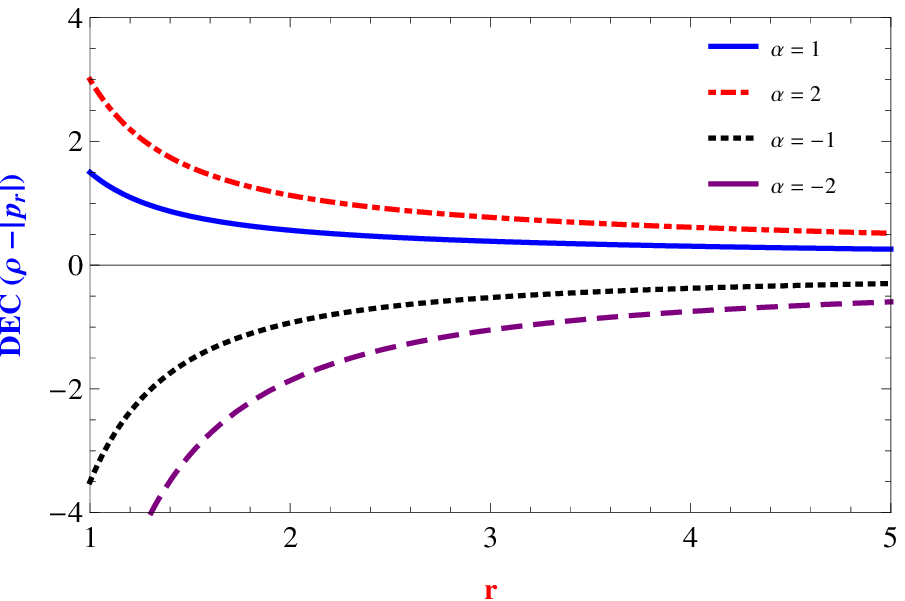}
	(b)\includegraphics[width=8cm, height=8cm, angle=0]{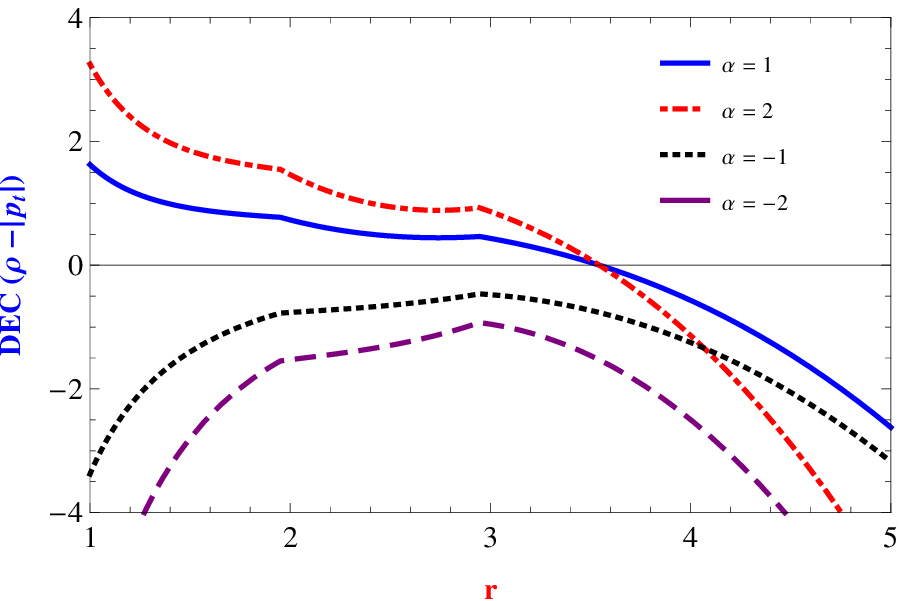}
	\caption {Variation of DECs ($\rho -|p_r|, \,\rho-|p_t|$) for throat radius $r_0 = 1$.}
\end{figure}

\begin{figure}
	(a)\includegraphics[width=8cm, height=8cm, angle=0]{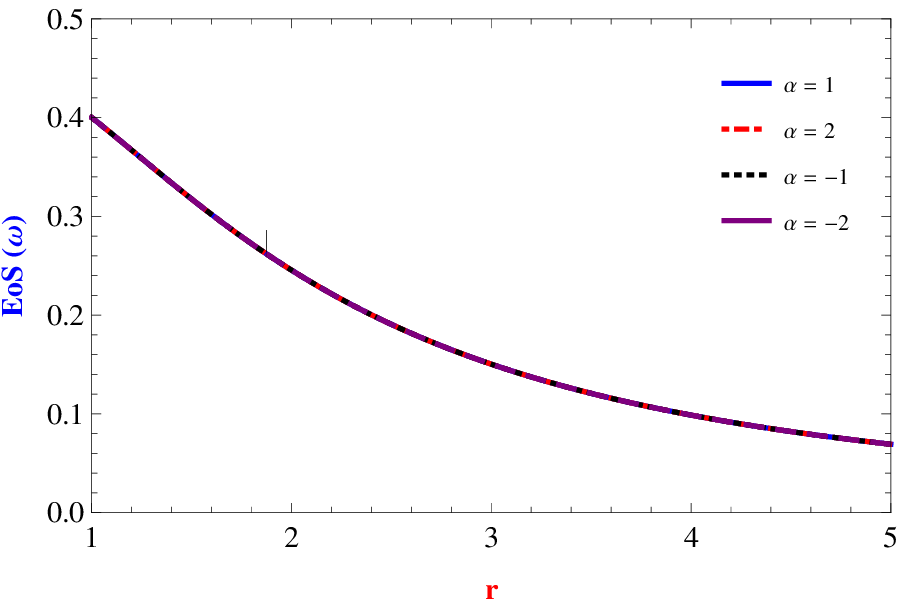}
	(b)\includegraphics[width=8cm, height=8cm, angle=0]{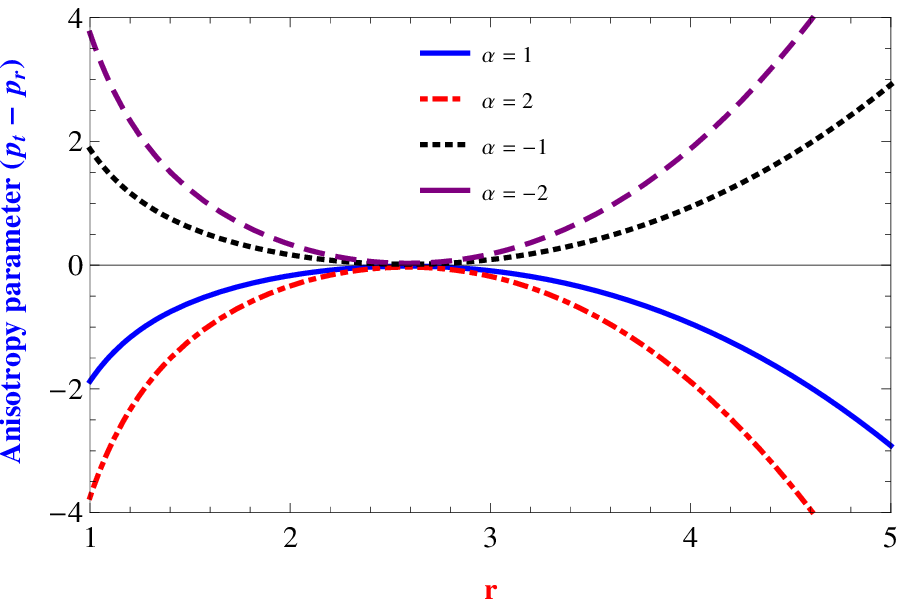}
	\caption {Variation of EoS parameter ($\omega$) and anisotropy parameter ($p_t-p_r$) for throat radius $r_0 = 1$.}
\end{figure}

\subsection{Wormhole solutions for non-linear form of $f(Q)$ i.e. $f(Q)=\alpha Q^2 + \beta$}

This section comprises the case where we  study the wormhole space time for non-linear function $f(Q) =\alpha Q^2 + \beta$ or the power law form of $f(Q)$  where $\alpha$ and $\beta$ are the constants. The power law model has already approved the radiation and CDM dominated background of the Universe. The shape function and the redshift functions are the same as we have used in the above linear case. Solving the field equations for this case, the components come out to be as follows

\begin{eqnarray}\label{eq23}
	\rho& =&\frac{1}{2{r}^{6}
		\left( \ln  \left( 2 \right)  \right) ^{2} \left( r+1 \right) ^{2}},\left\lbrace \left( 16\,r-56+8\,{r}^{2}+32\,{r}^{3} \right)  \left( 
	\ln  \left( r+1 \right)  \right) ^{2}-64\, \left( -1+r \right) 
	\left( -1/4\right.\right.\nonumber\\
	&+&
	\left.\left. \left( {r}^{2}+5/4\,r+7/4 \right) \ln  \left( 2 \right) 
	\right) r\ln  \left( r+1 \right) +\ln  \left( 2 \right)  \left( 
	\left(2\,\beta {r}^{5}+32\,{r}^{3}+\beta {r}^{6}+16\,r+\beta {r}^{4}\right.\right.\right.\nonumber\\
	&-&
	\left.\left.\left.56+8\,{r}^{2}
	\right) \ln  \left( 2 \right) +16-16\,r \right) {r}^{2}\right\rbrace
\end{eqnarray}

\begin{eqnarray}\label{eq24}
	p_r &=& \frac{1}{2{r}^{6} \left( \ln  \left( 2
		\right)  \right) ^{2} \left( r+1 \right) ^{2} }\left\lbrace -24\, \left( -1+r \right) ^{2} \left( \ln  \left( r+1
	\right)  \right) ^{2}
	+32\,\ln  \left( 2 \right) r \left( -1+r
	\right)  \left( r-2 \right) \ln  \left( r+1 \right) \right.\nonumber\\
	&-&
	\left. \left( \ln 
	\left( 2 \right)  \right) ^{2}{r}^{2} \left( \beta {r}^{6}+2\,\beta {r}^{5}+\beta {r
	}^{4}+8\,{r}^{2}-48\,r+40 \right) \right\rbrace
\end{eqnarray}

\begin{eqnarray}\label{eq25}
	p_t&=&\frac{1}{2{r}^{6} \left( \ln  \left( 2 \right) 
		\right) ^{2} \left( r+1 \right) ^{3} },\left\lbrace \left( -8\,r+8+8\,{r}^{2}+8\,{r}^{3}-16\,{r}^{4}
	\right)  \left( \ln  \left( r+1 \right)  \right) ^{2}\right.\nonumber\\
	&+&
	\left. \left( 
	\left( -16\,r+32\,{r}^{5}-16\,{r}^{3} \right) \ln  \left( 2 \right) +
	16\,{r}^{2}-16\,{r}^{3} \right) \ln  \left( r+1 \right) -\ln  \left( 2
	\right)  \left(  \left( r+1 \right)  \left( \beta {r}^{6}\right. \right.\right.\nonumber\\
	&+&
	\left.\left.\left.	2\,\beta {r}^{5}+\beta {r}
	^{4}+16\,{r}^{3}-8\,{r}^{2}-8 \right) \ln  \left( 2 \right) +16\,r-16
	\,{r}^{2} \right) {r}^{2}\right\rbrack
\end{eqnarray}
Here also, various energy conditions are coined for positive as well as negative values of $\alpha$. The character of energy density can be pursued from Fig. 5(a) which comes out to be positive for positive values of $\alpha$. For $\alpha < 0$ energy density is negative throughout the region and approaches zero with $r$ going on. We mainly focus on the solutions of wormholes where the existence of exotic matter can be avoided. For negative $\alpha$ the energy density is negative. Also the NEC is violated for the negative values of $\alpha$ which suggests the presence of non-ordinary matter at the wormhole throat. Now as we are observing from Fig. 5(b) and 6(a), both the NECs are validated for $r \leq 3.54 $ for $\alpha > 0$. The strong energy condition is violated throughout the region for every value of $\alpha$, as can be seen from Fig. 6(b). The radial and tangential both DEC's are devised in Fig. 7(a) and (b). The radial DEC is violated for every $r \geq r_0$ while the tangential DEC is satisfied partially. The anisotropy parameter is positive everywhere for positive values of $\alpha$. Moving away from the throat $r_0 =  1$, the anisotropy parameter tends to zero and then again increases towards the positive side as $r$ increases further. The EoS parameter $\omega$ is drafted in Fig. 8(a) which is negative for every value of $\alpha$ and for all $r \geq r_0$. At throat $r_0 =1$ the value of $\omega$ is -1 also more generally for $r \geq r_0$, $ -1 \leq \omega <- 1/3$ which shows the presence of exotic matter around the throat which in turn also shows the violation of energy conditions at the wormhole throat by some arbitrary values for positive $\alpha$.

\begin{figure}
	(a)\includegraphics[width=8cm, height=8cm, angle=0]{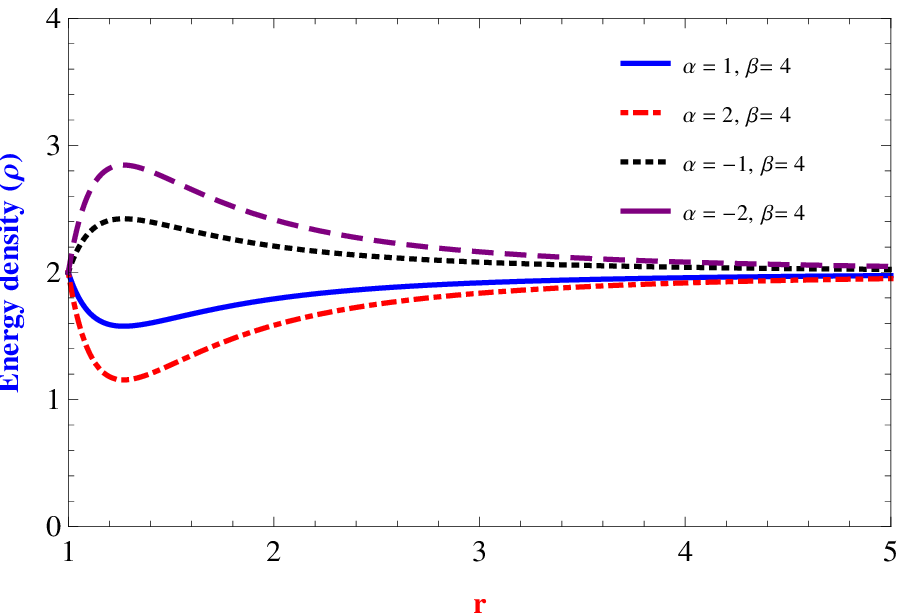}
	(b)\includegraphics[width=8cm, height=8cm, angle=0]{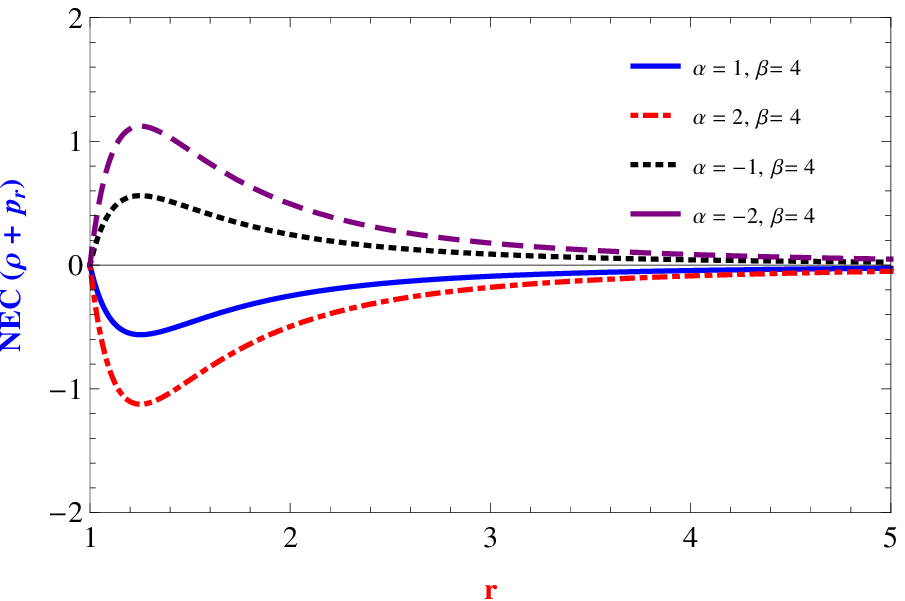}
	\caption {Variation of Energy density ($\rho$) and NEC ($\rho + p_r$) for throat radius $r_0 = 1$ }
\end{figure}
\begin{figure}
	(a)\includegraphics[width=8cm, height=8cm, angle=0]{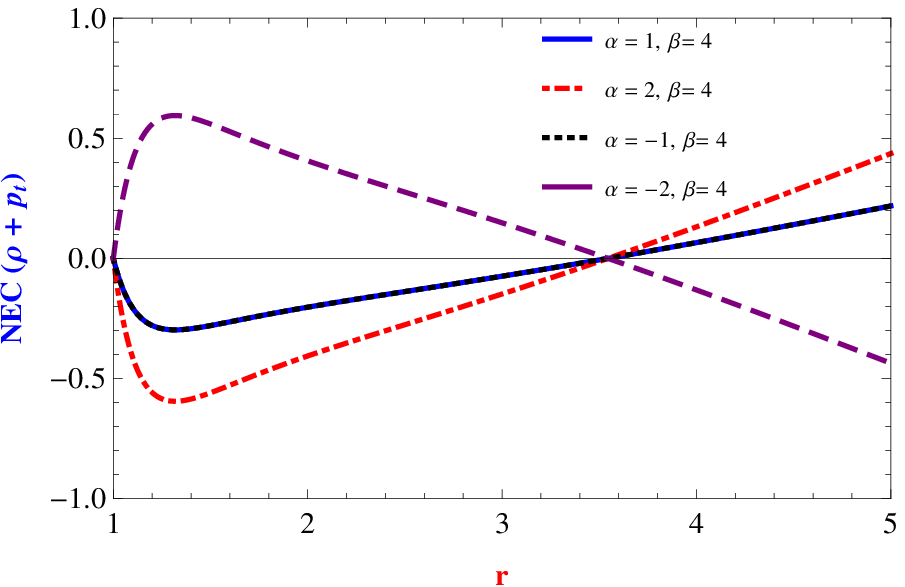}
	(b)\includegraphics[width=8cm, height=8cm, angle=0]{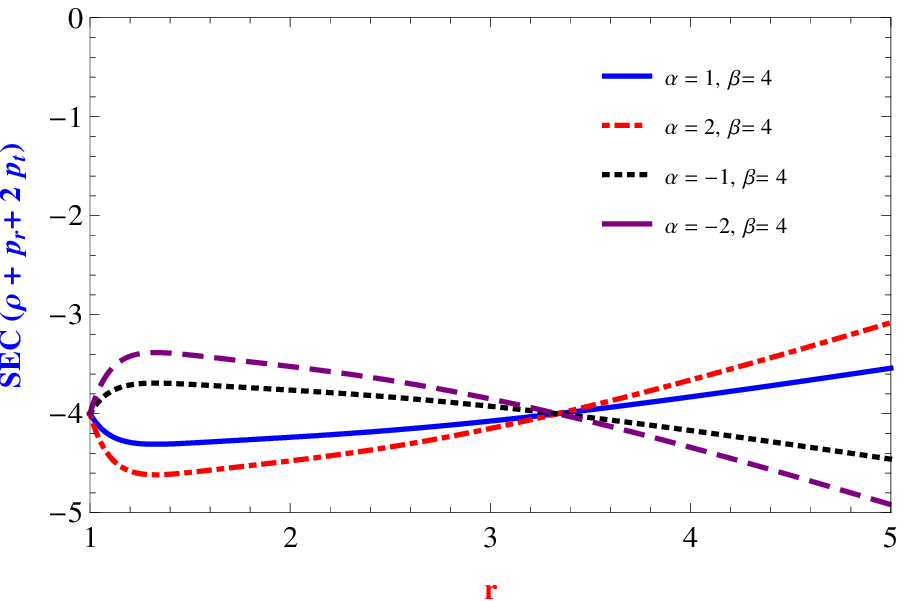}
	\caption {Variation of NEC ($\rho+p_t$) and  SEC ($\rho + p_r+ 2p_t$) for throat radius $r_0 = 1$.}
\end{figure}
\begin{figure}
	(a)\includegraphics[width=8cm, height=8cm, angle=0]{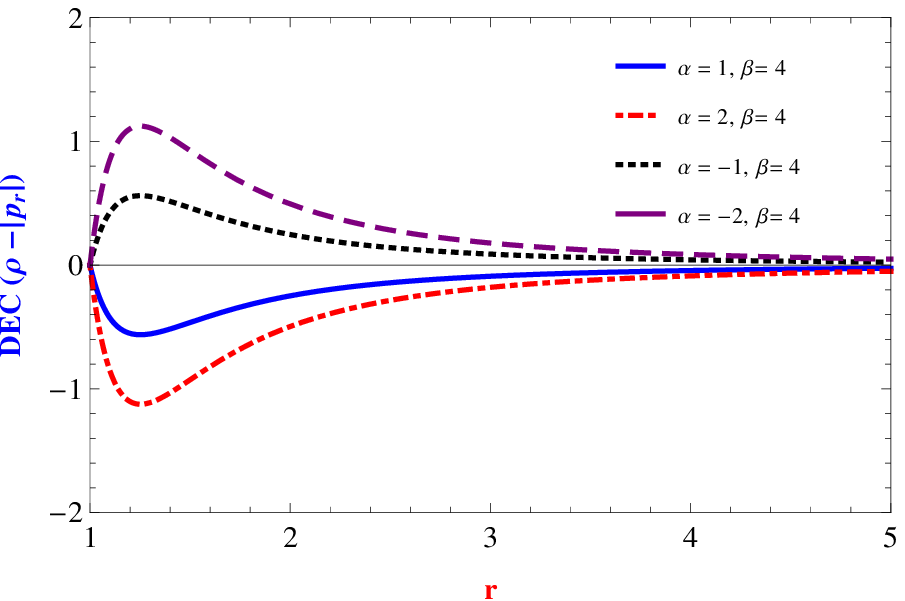}
	(b)\includegraphics[width=8cm, height=8cm, angle=0]{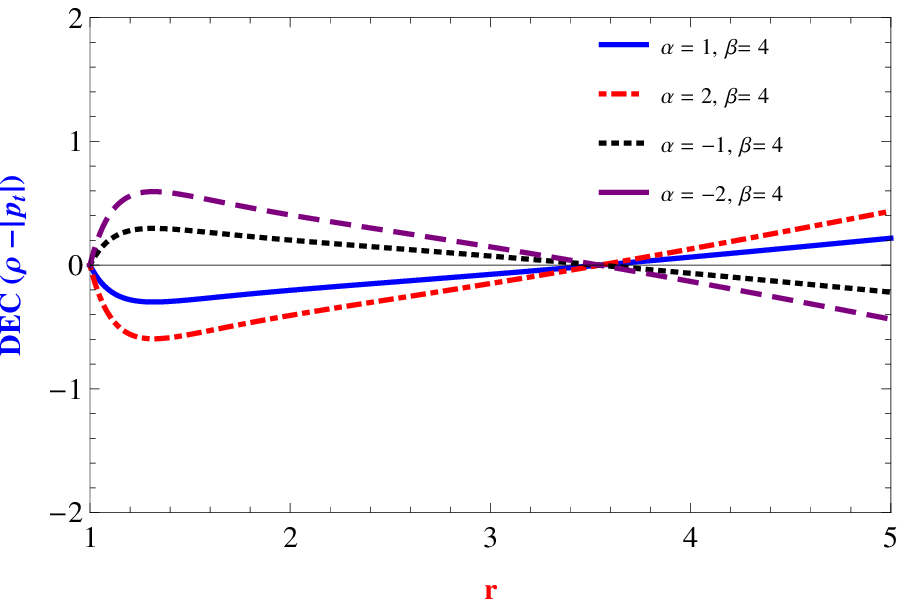}
	\caption {Variation of DECs ($\rho -|p_r|, \,\rho-|p_t|$) for throat radius $r_0 = 1$.}
\end{figure}

\begin{figure}
	(a)\includegraphics[width=8cm, height=8cm, angle=0]{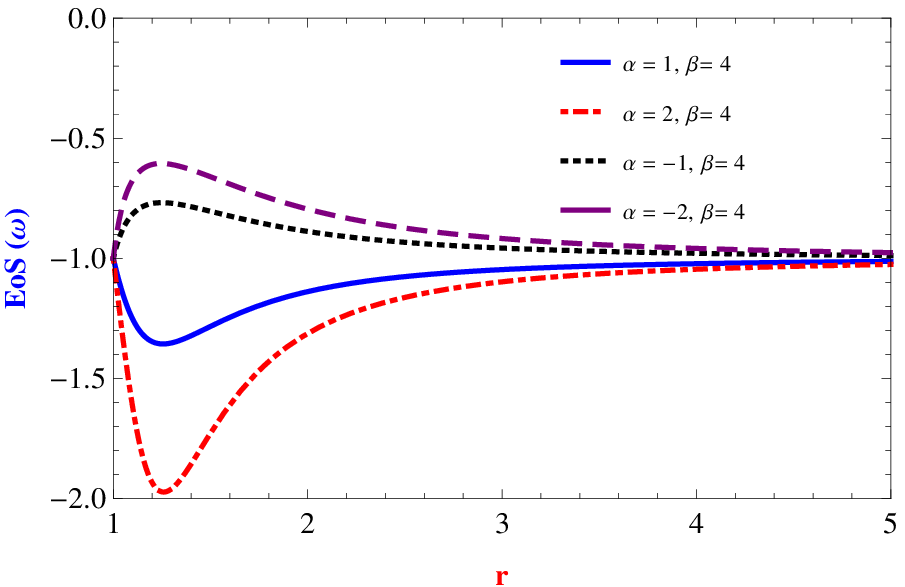}
	(b)\includegraphics[width=8cm, height=8cm, angle=0]{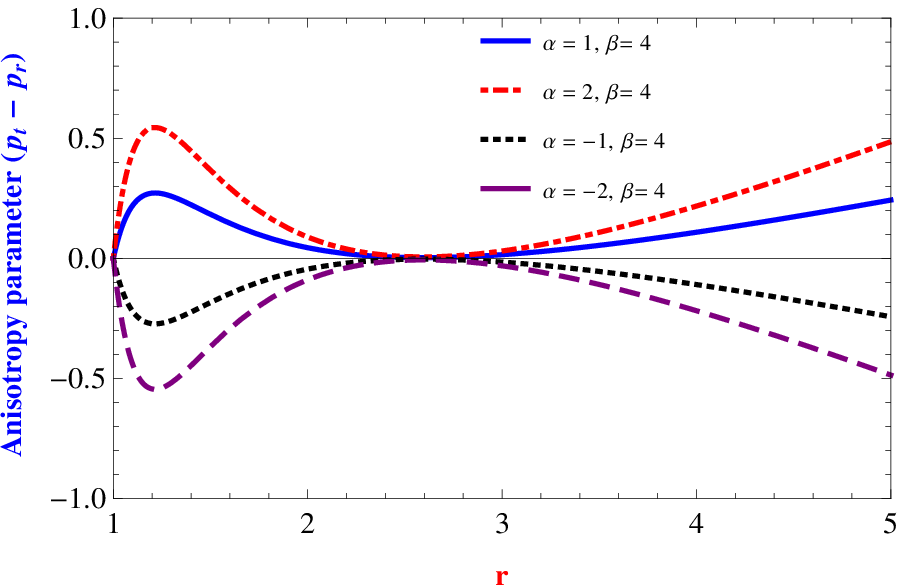}
	\caption {Variation of EoS parameter ($\omega$) and anisotropy parameter ($p_t-p_r$) for throat radius $r_0 = 1$.}
\end{figure}



\begin{figure}
	(a)\includegraphics[width=8cm, height=8cm, angle=0]{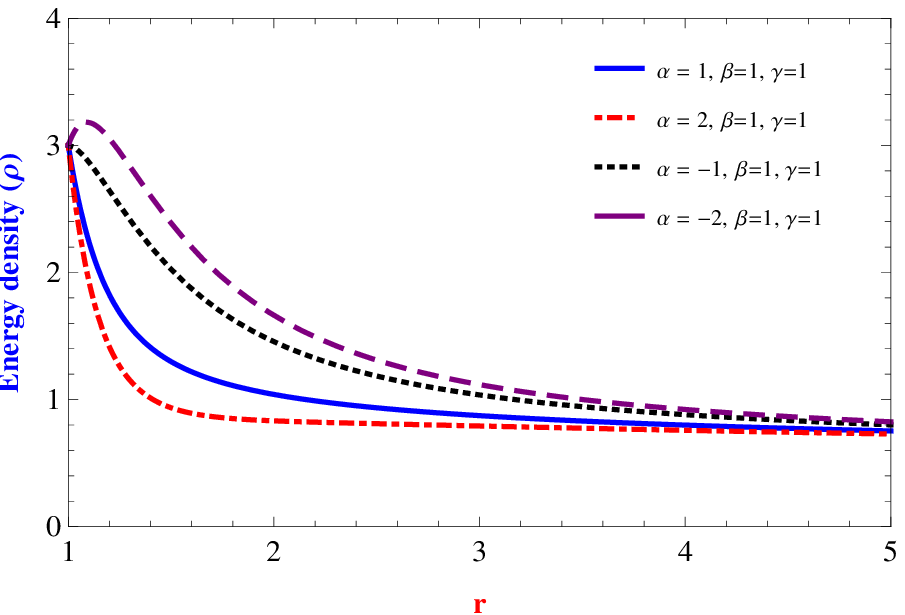}
	(b)\includegraphics[width=8cm, height=8cm, angle=0]{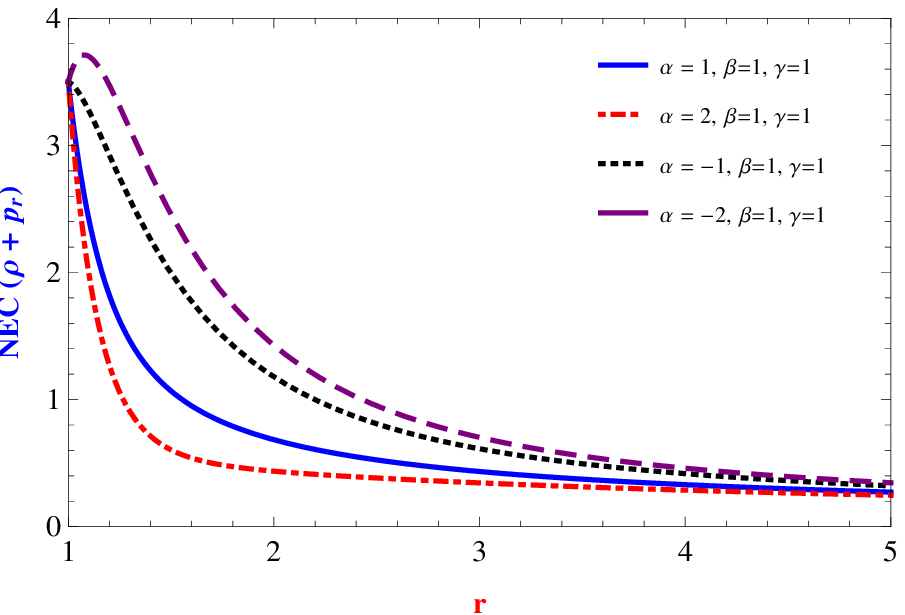}
	\caption {Variation of Energy density ($\rho$) and NEC ($\rho + p_r$) for throat radius $r_0 = 1$ }
\end{figure}
\begin{figure}
	(a)\includegraphics[width=8cm, height=8cm, angle=0]{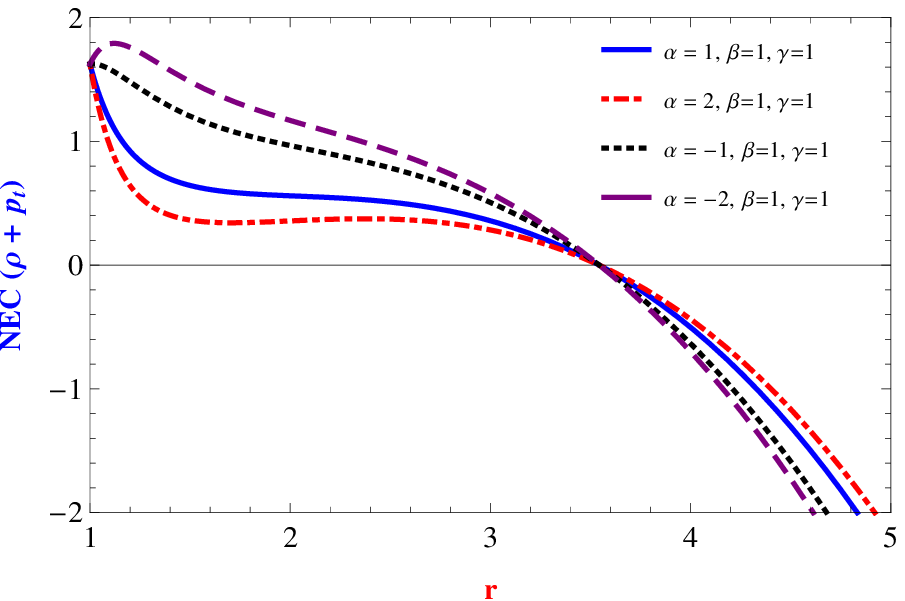}
	(b)\includegraphics[width=8cm, height=8cm, angle=0]{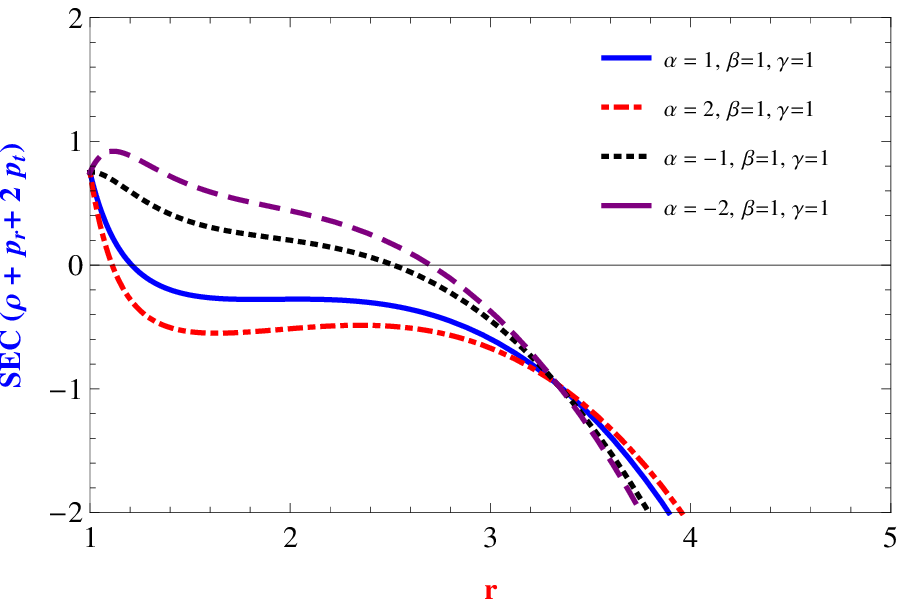}
	\caption {Variation of NEC ($\rho+p_t$) and  SEC ($\rho + p_r+ 2p_t$) for throat radius $r_0 = 1$.}
\end{figure}
\begin{figure}
	(a)\includegraphics[width=8cm, height=8cm, angle=0]{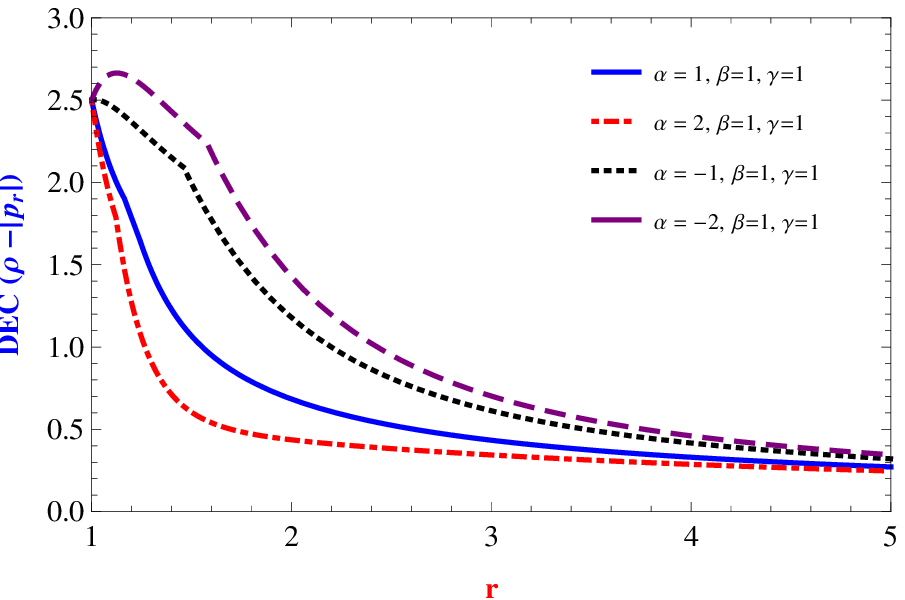}
	(b)\includegraphics[width=8cm, height=8cm, angle=0]{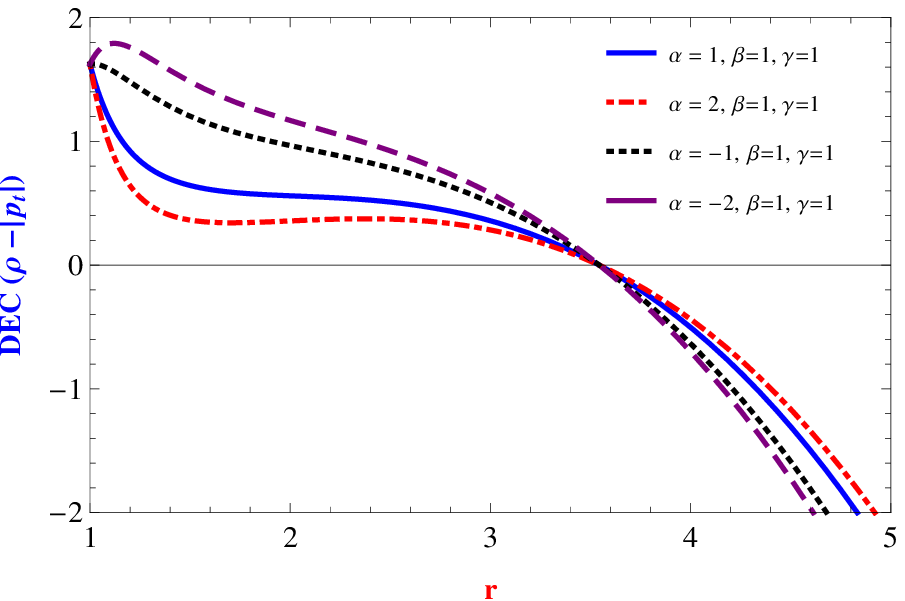}
	\caption {Variation of DECs ($\rho -|p_r|, \,\rho-|p_t|$) for throat radius $r_0 = 1$.}
\end{figure}

\begin{figure}
	(a)\includegraphics[width=8cm, height=8cm, angle=0]{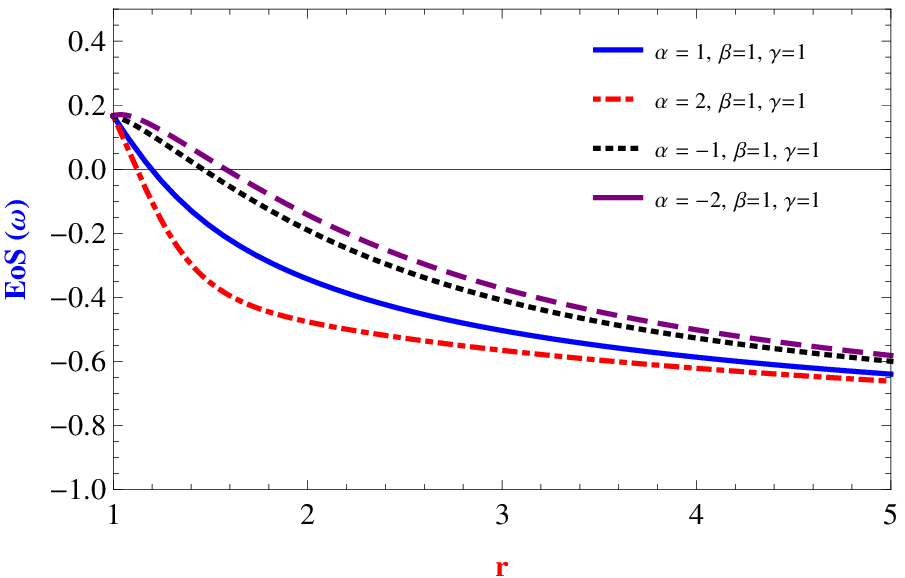}
	(b)\includegraphics[width=8cm, height=8cm, angle=0]{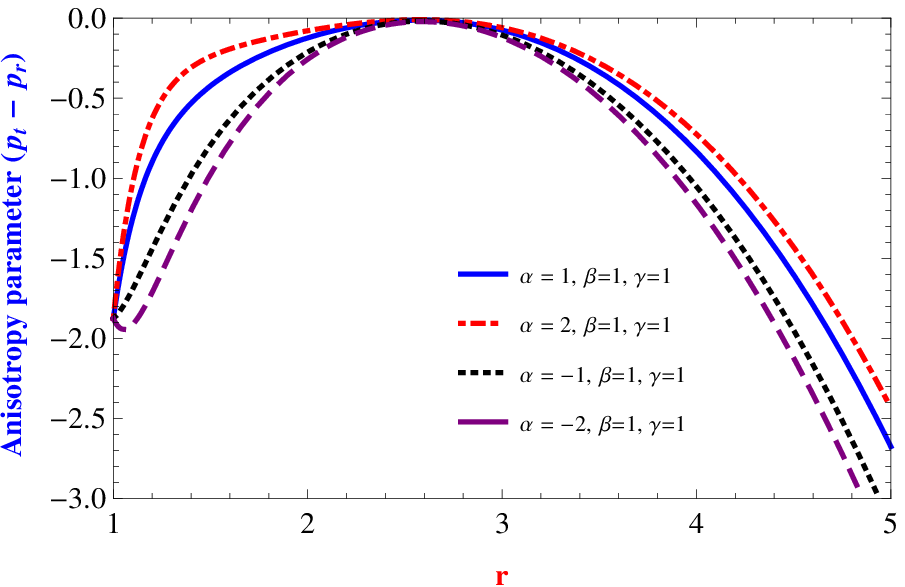}
	\caption {Variation of EoS parameter ($\omega$) and anisotropy parameter ($p_t-p_r$) for throat radius $r_0 = 1$.}
\end{figure}


\subsection{Wormhole solutions for quadratic form of $f(Q)$ i.e. $f(Q)=\alpha Q^2 + \beta Q + \gamma$}

Here in this section, we take a more general quadratic form of function $f(Q)$. The above two cases of linear and non -linear $f(Q)$  have been examined with different shape functions for possible wormhole geometry \cite{ref28}. But we  try to get an insight into the wormhole geometry for this quadratic gravity $f(Q)$. As we have already mentioned, the shape function $b(r)$ and the redshift function  $\phi(r)$ are same as the previous cases. Using this quadratic model, the field equations (\ref{eq12}), (\ref{eq13}) and (\ref{eq14}) can be developed to give the stress energy components as

\begin{eqnarray}\label{eq26}
	\rho&=& \frac{1}{2 r^6 (r+1) (r+{r_0})^2}\left\lbrace-4 \alpha  (r+1) {r_0}^2 \left(4 r^3-3 r^2+2 r {r_0} (3-2 {r_0})-3 {r_0}^2\right) \log ^2\left(\frac{r+1}{{r_0}+1}\right)\right.\nonumber\\
	&-&
	\left. 4 r {r_0} \log \left(\frac{r+1}{{r_0}+1}\right) \left(\beta  r^6+2 \beta  r^5 {r_0}+r^4 \left(\beta  \left({r_0}^2+2 {r_0}-1\right)-8 \alpha \right)-2 r^3 \left(\alpha -\beta  {r_0}^2\right)\right.\right.\nonumber\\
	&+&
	\left.\left. r^2 \left(2 \alpha  \left(4 {r_0}^2-7 {r_0}+3\right)+\beta  {r_0}^2\right)+2 \alpha  r {r_0} (7 {r_0}-6)+2 \alpha  {r_0}^2 ({r_0}+3)\right)\right.\nonumber\\
	&+&
	\left. r^2 \left(\gamma  r^7+r^6 (4 \beta +\gamma +2 \gamma  {r_0})+r^5 {r_0} (8 \beta +\gamma  ({r_0}+2))+r^4 \left(-16 \alpha +2 \beta  \left(2 {r_0}^2+5 {r_0}-2\right)+\gamma  {r_0}^2\right)\right.\right.\nonumber\\
	&-&
	\left.\left. 4 r^3 \left(\alpha -3 \beta  {r_0}^2\right)+2 r^2 \left(2 \alpha  \left(4 {r_0}^2-8 {r_0}+3\right)+\beta  ({r_0}+2) {r_0}^2\right)+4 \alpha  r {r_0} (7 {r_0}-6)\right.\right.\nonumber\\
	&+&
	\left.\left. 4 \alpha  {r_0}^2 (2 {r_0}+3)\right)\right\rbrace,
\end{eqnarray}

\begin{eqnarray}\label{eq27}
	p_r &=& \frac{1}{2 r^6 (r+{r_0})^2}\left\lbrace2 r {r_0} (r-{r_0}) \log \left(\frac{r+1}{{r_0}+1}\right) \left(\beta  r^3+\beta  r^2 {r_0}-8 \alpha  r+16 \alpha  {r_0}\right)\right.\nonumber\\
	&-&
	\left. r^2 \left(\gamma  r^6+2 \gamma  r^5 {r_0}+\gamma  r^4 {r_0}^2-4 \beta  r^3 {r_0}-4 r^2 \left(\alpha +\beta  {r_0}^2\right)+24 \alpha  r {r_0}-20 \alpha  {r_0}^2\right)\right.\nonumber\\
	&+&
	\left. 12 \alpha  {r_0}^2 (r-{r_0})^2 \log ^2\left(\frac{r+1}{{r_0}+1}\right)\right\rbrace,
\end{eqnarray}

\begin{eqnarray}\label{eq28}
	p_t&=& \frac{1}{2 r^6 (r+1) (r+{r_0})^2}\left\lbrace-4 \alpha  (r+1) {r_0}^2 \left(r^7-r^6 {r_0}-2 r^3+r^2 (2 {r_0}-1)+{r_0}^2\right) \log ^2\left(\frac{r+1}{{r_0}+1}\right)\right.\nonumber\\
	&+&
	\left. r^2 \left(\beta  r^9 {r_0}+\beta  r^8 {r_0}^2-r^7 (\gamma +4 \alpha  {r_0})+r^6 \left(-2 \beta -\gamma +4 \alpha  {r_0}^2-2 \gamma  {r_0}\right)-r^5 (2 \beta  ({r_0}+1)\right.\right.\nonumber\\
	&+&
	\left.\left.\gamma  {r_0} ({r_0}+2))+r^4 (8 \alpha -{r_0} (4 \beta +\gamma  {r_0}))-2 r^3 (\alpha  (4 {r_0}-6)+\beta  {r_0} ({r_0}+1))\right.\right.\nonumber\\
	&-&
	\left.\left. 2 r^2 \left(\beta  {r_0}^2+\alpha  (4 {r_0}-2)\right)-4 \alpha  r {r_0}^2-4 \alpha  {r_0}^2\right)+r {r_0} \log \left(\frac{r+1}{{r_0}+1}\right) \left(-\beta  r^{10}-\beta  r^9 ({r_0}+1)\right.\right.\nonumber\\
	&+&
	\left.\left. r^8 (4 \alpha -\beta  {r_0})+4 \alpha  r^7+r^6 \left(2 \beta -4 \alpha  {r_0}^2-4 \alpha  {r_0}\right)+2 \beta  r^5 ({r_0}+1)+4 r^4 (\beta  {r_0}-4 \alpha )\right.\right.\nonumber\\
	&+&
	\left.\left. 2 r^3 (4 \alpha  (2 {r_0}-3)+\beta  {r_0} ({r_0}+1))+2 r^2 \left(\beta  {r_0}^2+\alpha  (8 {r_0}-4)\right)+8 \alpha  r {r_0}^2+8 \alpha  {r_0}^2\right)\right\rbrace.
\end{eqnarray}

Fig. 9(a) depicts the energy density $\rho$ against both negative and positive values of constant $\alpha$. As we can interpret from the figure, energy density is positive at the throat and everywhere for $r\geq r_0$. Here we have fixed the values of other constants as $\beta=1$ and $\gamma=1$. It is very interesting to see in Figs. 9(b) and 10(a), that both the NEC's are satisfied at the throat and beyond as $r$ increases. The tangential NEC, however, is not satisfied for $r > 3.53$. This validation implies that for around the throat the WEC is authenticated for all values of $\alpha$. From Fig. 10(b) it is clear that SEC is validated at the throat and in a small region  beyond the throat for $r \geq r_0$. The radial and tangential DECs are plotted  in Fig. 11(a) and (b). It is gathered from the figure that the radial DEC is satisfied everywhere for $r \geq r_0$ while the tangential DEC is validated at the throat and beyond for $r \leq 3.53 $ and violates as $r$ increases further. As far as the anisotropy parameter is concerned, it is drawn in Fig. 12(b) and is negative everywhere for $r \geq r_0$ except for some values of $r$ where it reaches to zero and again becomes negative for larger values of $r$. This negative value of the anisotropy parameter shows that the geometry around the throat has a repulsive nature. The presence of ordinary or non-exotic matter near the throat is found in Fig. 12(a) i.e. $\omega >-1/3$ for all values of $\alpha$. But it is also suggesting the existence of quintessence $-1 <\omega<-1/3$ as $r$ goes higher i.e. at the region away from the throat.

\section{Discussion and Conclusion}

The solutions of wormholes in GR are restricted to the violation of averaged NEC which is an inevitable consequence in GR which leads to the presence of exotic matter. However in modified gravity it is tried to avoid the exotic form of matter by  considering the higher order terms which support the wormhole  geometry. In our  present work, our main objective is to explore the wormhole solution with standard fluid.  We have examined the spherically symmetric and static traversable WH in the backdrop of $f(Q)$ gravity i. e. the symmetric tele-parallel gravity where the term $Q$ is the non-metricity term describing the gravitational interactions. Many researchers are nowadays interested in exploring the cosmological aspects and  wormhole solutions in the framework of $f(Q)$ gravity. As stated earlier in this section, in GR to have a traversable wormhole, the violation of NEC is a must which signifies the exotic matter in the wormhole throat. To avoid this situation and to find a more realistic solution, We have taken three functional forms of $f(Q)$. The motivation of our work came from  in which Hassan et al. \cite{ref28}  have explored the wormhole solutions taking linear form $f(Q)=\alpha Q$ and the non-linear form $f(Q)=\alpha Q^2 + \beta$. However, they found  an appropriate solution for traversable wormhole geometry in $f(Q)$ which violates the NEC at the throat and indicates the presence of exotic matter. Though they measured the quantity of exotic matter to be small. In this work, we have examined the shape function  $b(r)={\frac{r_{0}\ln(r+1)}{\ln({r_0}+1)}}$ for all the three cases and have considered the same variable redshift function $\phi(r)=\ln  \left( {\frac {r_{{0}}}{r}}+1 \right) $ throughout the study. This shape function is already
examined for the possible solution in other modified gravities such as $f(R)$ and 4-D EGB theory \cite{ref21a,ref31}, hence, it has already been proved there to have satisfied all the necessary conditions of shape function.\\

First, the possibility of a traversable wormhole for a linear form of $f(Q)$ is examined by studying all the energy conditions against the radial coordinate $r$ and for negative as well as positive values of $\alpha$. The radius of the throat is $r_0  = 1$. We can collect from the figures that for positive values of $\alpha$, the NEC and WEC are satisfied near the throat which implies the presence of normal matter in the throat. Both the DEC's  are also satisfied here. If we look at the values of the EoS parameter $\omega$, it is clear that there is no presence of exotic matter as its value is greater than $-1/3$. These outcomes imply that the wormhole solutions obtained here are justifiable in the $f(Q)$ gravity.\\

In the second case, the wormhole geometries are discussed in the backdrop of a non-linear Lagrangian $f(Q)$. To explore the possibility of a non exotic traversable wormhole, all the energy conditions are examined. As in the case of linear $f(Q)$, For positive $\alpha$ the NEC profiles are validated in the vicinity of the throat. The radial DEC is satisfied throughout while the lateral DEC is satisfied in the neighborhood of the throat for $r \geq r_0=1$ and then as $r$ increases it becomes negative. These profiles also suggest the existence of wormhole solutions with non-exotic matter. Although the EoS parameter in this case gives an indication towards the presence of exotic matter near the throat. Though it has been shown in previous works within the framework of other modified gravity that the presence of exotic matter can be minimized in the throat \cite{ref17}.\\

The third case we have considered is the case of a more general quadratic form of $f(Q)$ with constants $\alpha$ and $\gamma$. Again the same shape function is tried for a plausible solution of wormholes. If we go through the various energy conditions, we are amazed to see that this quadratic Lagrangian  has very interesting results. As we can see the NEC and WEC are satisfied for all values of alpha near the throat which gives an attractive nature of fluid in the throat. Both the DEC's are also justified around the throat. The energy conditions are also consistent with the EoS parameter $\omega$ since at the throat the value of $\omega >-1/3$ which shows that at the throat only normal matter is present. Hence this case gives a feasible solution of traversable wormhole without need of exotic matter.\\

Hence, we can conclude that with an appropriate shape function and redshift function, we can find plausible solutions which support the traversable wormhole geometry within the framework of $f(Q)$ gravity. Also with particular forms of $f(Q)$ the existence of exotic matter can be minimized and even can be avoided completely in some cases. 




\begin{thebibliography}{99}

\bibitem{ref1}
K.~Akiyama \textit{et al.} [Event Horizon Telescope],
``First M87 Event Horizon Telescope Results. IV. Imaging the Central Supermassive Black Hole,''
Astrophys. J. Lett. \textbf{875} (2019) no.1, L4

\bibitem{ref2}
L.~Blackburn, S.~Doeleman, J.~Dexter, J.~L.~G\'omez, M.~D.~Johnson, D.~C.~Palumbo, J.~Weintroub, K.~L.~Bouman, A.~A.~Chael and J.~R.~Farah, \textit{et al.}
``Studying Black Holes on Horizon Scales with VLBI Ground Arrays,''
[arXiv:1909.01411 [astro-ph.IM]]

\bibitem{ref3}
R.~Abuter \textit{et al.} [GRAVITY],
``Detection of the Schwarzschild precession in the orbit of the star S2 near the Galactic centre massive black hole,''
Astron. Astrophys. \textbf{636} (2020), L5

\bibitem{ref4}
E.~Gourgoulhon, A.~Le Tiec, F.~H.~Vincent and N.~Warburton,
``Gravitational waves from bodies orbiting the Galactic Center black hole and their detectability by LISA,''
Astron. Astrophys. \textbf{627} (2019), A92
	
\bibitem {ref5}	
	L . Flamm, ``Comments on Einstein’s theory of gravity,''  Phys. Z. \textbf{17} (1916), 448 
	
\bibitem {ref6}      	
	M. Visser, Lorentzian wormholes: From Einstein to Hawking, Springer-Verlag, Berlin, (1997)

\bibitem {ref7}
 A. Einstein, N. Rogen, ``The particle problem in the general theory of relativity,''  Phys. Rev. \textbf{48} (1935), 73 

\bibitem{ref8}
S.~Capozziello and M.~De Laurentis,
``Extended Theories of Gravity,''
Phys. Rept. \textbf{509} (2011), 167-321
 
 \bibitem{ref9}
 K.~Jusufi and A.~\"Ovg\"un,
 ``Gravitational Lensing by Rotating Wormholes,''
 Phys. Rev. D \textbf{97} (2018) no.2, 024042
 
 \bibitem{ref10}
 A.~\"Ovg\"un, K.~Jusufi and \.I.~Sakall\i{},
 ``Exact traversable wormhole solution in bumblebee gravity,''
 Phys. Rev. D \textbf{99} (2019) no.2, 024042
 
 \bibitem{ref11}
   M.~S.~Morris and K.~S.~Thorne,
   ``Wormholes in space-time and their use for interstellar travel: A tool for teaching general relativity,''
  Am. J. Phys. \textbf{56} (1988), 395-412
  
  \bibitem{ref12} 
     G. Clement,
  ``The Ellis geometry,''
     Am. J. Phys. \textbf{57}(11) (1989), 967
     
     \bibitem{ref13}
       H.~a.~Shinkai and S.~A.~Hayward,
      ``Fate of the first traversible wormhole: Black hole collapse or inflationary expansion,''
      Phys. Rev. D \textbf{66} (2002), 044005
      
      \bibitem{ref14}   
        M.~S.~Morris, K.~S.~Thorne and U.~Yurtsever,
        ``Wormholes, Time Machines, and the Weak Energy Condition,''
        Phys. Rev. Lett. \textbf{61} (1988), 1446-1449
          
        \bibitem{ref15}
        S.~W.~Kim and K.~S.~Thorne,
        ``Do vacuum fluctuations prevent the creation of closed timelike curves?,''
        Phys. Rev. D \textbf{43} (1991), 3929-3947
          
       \bibitem{ref16}
         M.~Visser,
         ``From wormhole to time machine: Comments on Hawking's chronology protection conjecture,''
         Phys. Rev. D \textbf{47} (1993), 554-565
         
         \bibitem{ref17}
           K.~Jusufi, A.~Banerjee and S.~G.~Ghosh,
           ``Wormholes in 4D Einstein\textendash{}Gauss\textendash{}Bonnet gravity,''
           Eur. Phys. J. C \textbf{80} (2020) no.8, 698
      
        \bibitem{ref18}
        S.~V.~Sushkov,
        ``Wormholes supported by a phantom energy,''
        Phys. Rev. D \textbf{71} (2005), 043520 
        
        \bibitem{ref19}
        D.~Wang and X.~H.~Meng,
        ``Wormholes supported by phantom energy from Shan\textendash{}Chen cosmological fluids,''
        Eur. Phys. J. C \textbf{76} (2016) no.3, 171  
        
        \bibitem{ref20}
        P.~K.~Sahoo, P.~H.~R.~S.~Moraes, P.~Sahoo and G.~Ribeiro,
        ``Phantom fluid supporting traversable wormholes in alternative gravity with extra material terms,''
        Int. J. Mod. Phys. D \textbf{27} (2018) no.16, 1950004
        
        \bibitem{ref21}
        F.~S.~N.~Lobo,
        ``Phantom energy traversable wormholes,''
        Phys. Rev. D \textbf{71} (2005), 084011
        
        \bibitem{ref21a}
        Shweta, A.~K.~Mishra and U.~K.~Sharma,
        ``Traversable wormhole modelling with exponential and hyperbolic shape functions in $F(R,T)$ framework,''
        Int. J. Mod. Phys. A \textbf{35} (2020) no.25, 2050149 
        
        \bibitem{ref22}
        P.~H.~R.~S.~Moraes and P.~K.~Sahoo,
        ``Modelling wormholes in $f(R,T)$ gravity,''
        Phys. Rev. D \textbf{96} (2017) no.4, 044038
        
        
        \bibitem{ref22a}
        P.~K.~Sahoo, P.~H.~R.~S.~Moraes and P.~Sahoo,
        ``Wormholes in $R^2$ -gravity within the $f(R, T)$ formalism,''
        Eur. Phys. J. C \textbf{78} (2018) no.1, 46
        
        \bibitem{ref22b}
        U.~K.~Sharma and A.~K.~Mishra,
        ``Wormholes Within the Framework of $f(R, T)=R+\alpha R^2+\lambda T$ Gravity,''
        Found. Phys. \textbf{51} (2021) no.2, 50
        \bibitem{ref22c}
        M.~Zubair, S.~Waheed, G.~Mustafa and H.~Ur Rehman,
        ``Noncommutative inspired wormholes admitting conformal motion involving minimal coupling,''
        Int. J. Mod. Phys. D \textbf{28} (2019) no.04, 1950067
        \bibitem{ref22d}
        A. K. Mishra and U. K.  Sharma, ``Wormhole models in $R^2$-gravity for $f(R, T)$ theory with a hybrid shape function,'' 	Can. J. Phys. \textbf{99}(6) (2021) 481
        
        \bibitem{ref22e}
        M.~F.~Shamir, G.~Mustafa and A.~Fazal,
        ``Non-commutative wormhole solutions in exponential gravity with matter coupling,''
        New Astron. \textbf{83} (2021), 101459; A.~K.~Mishra, V.~C.~Dubey and U.~K.~Sharma,
        ``Two different shape functions for wormholes in $f(R)$ theory with non-commutative geometry and Lorentzian distribution,''
        Int. J. Geom. Meth. Mod. Phys. \textbf{17} (2020) no.11, 2050155; A.~K.~Mishra and U.~K.~Sharma,
        ``A new shape function for wormholes in $f(R)$ gravity and General Relativity,''
        New Astron. \textbf{88} (2021), 101628
        
        \bibitem{ref22f}
        G.~Mustafa, X.~Tie-Cheng, M.~Ahmad and M.~F.~Shamir,
        ``Anisotropic spheres via embedding approach in R+\ensuremath{\beta}R2 gravity with matter coupling,''
        Phys. Dark Univ. \textbf{31} (2021), 100747
        
        \bibitem{ref22g}
        P.~H.~R.~S.~Moraes and P.~K.~Sahoo,
        ``Wormholes in exponential $f(R,T)$ gravity,''
        Eur. Phys. J. C \textbf{79} (2019) no.8, 677
        
        \bibitem{ref23}
        S.~Capozziello, T.~Harko, T.~S.~Koivisto, F.~S.~N.~Lobo and G.~J.~Olmo,
        ``Wormholes supported by hybrid metric-Palatini gravity,''
        Phys. Rev. D \textbf{86} (2012), 127504
        
    \bibitem{ref23a}
    V.~De Falco, E.~Battista, S.~Capozziello and M.~De Laurentis,
    ``Reconstructing wormhole solutions in curvature based Extended Theories of Gravity,''
    Eur. Phys. J. C \textbf{81} (2021) no.2, 157  
    
    \bibitem{ref23b}
    G.~Mustafa, M.~Ahmad, A.~\"Ovg\"un, M.~F.~Shamir and I.~Hussain,
    ``Traversable wormholes in the extended teleparallel theory of gravity with matter coupling,''
    [arXiv:2104.13760 [gr-qc]].
    
    \bibitem{ref23c}
    K.~N.~Singh, A.~Banerjee, F.~Rahaman and M.~K.~Jasim,
    ``Conformally symmetric traversable wormholes in modified teleparallel gravity,''
    Phys. Rev. D \textbf{101} (2020) no.8, 084012  
    
    \bibitem{ref23d}
    C.~G.~Boehmer, T.~Harko and F.~S.~N.~Lobo,
    ``Wormhole geometries in modified teleparralel gravity and the energy conditions,''
    Phys. Rev. D \textbf{85} (2012), 044033
    
        
        
       \bibitem{ref24}
       R.~Aldrovandi and J.~G.~Pereira,
       ``Teleparallel Gravity: An Introduction,''
       doi:10.1007/978-94-007-5143-9
       
       
        \bibitem{ref27}
       J.~Beltr\'an Jim\'enez, L.~Heisenberg and T.~Koivisto,
       ``Coincident General Relativity,''
       Phys. Rev. D \textbf{98} (2018) no.4, 044048
       
   \bibitem{ref28e}
   R.~Lazkoz, F.~S.~N.~Lobo, M.~Ortiz-Ba\~nos and V.~Salzano,
   ``Observational constraints of $f(Q)$ gravity,''
   Phys. Rev. D \textbf{100} (2019) no.10, 104027
   \bibitem{ref28f}
   I.~Ayuso, R.~Lazkoz and V.~Salzano,
   ``Observational constraints on cosmological solutions of $f(Q)$ theories,''
   Phys. Rev. D \textbf{103} (2021) no.6, 063505
   
   \bibitem{ref28g}
   F.~K.~Anagnostopoulos, S.~Basilakos and E.~N.~Saridakis,
   ``First evidence that non-metricity $f(Q)$ gravity can challenge $\Lambda$CDM,''
   [arXiv:2104.15123 [gr-qc]].
   
   
   \bibitem{ref28h}
   S.~Mandal, P.~K.~Sahoo and J.~R.~L.~Santos,
   ``Energy conditions in $f(Q)$ gravity,''
   Phys. Rev. D \textbf{102} (2020) no.2, 024057
   
   \bibitem{ref28i}
   S.~Mandal, D.~Wang and P.~K.~Sahoo,
   ``Cosmography in $f(Q)$ gravity,''
   Phys. Rev. D \textbf{102} (2020), 124029
   
   \bibitem{ref28j}
   T.~Harko, T.~S.~Koivisto, F.~S.~N.~Lobo, G.~J.~Olmo and D.~Rubiera-Garcia,
   ``Coupling matter in modified $Q$ gravity,''
   Phys. Rev. D \textbf{98} (2018) no.8, 084043    
        
       \bibitem{ref28}
     Z.~Hassan, S.~Mandal and P.~K.~Sahoo,
       ``Traversable wormhole geometries in $f(Q)$ gravity,''
       Fortsch. Phys. \textbf{69} (2021), 2100023
     \bibitem{ref28a1}   
    Z.~Hassan, G.  Mustafa, P. K.  Sahoo, ``Wormhole Solutions in Symmetric Teleparallel Gravity with Noncommutative Geometry,'' Symmetry  {\bf13} (2021), 1260.    
       
       \bibitem{ref25}
       J.~W.~Maluf,
       ``The teleparallel equivalent of general relativity,''
       Annalen Phys. \textbf{525} (2013), 339-357
       
       \bibitem{ref26}
       V.~C.~de Andrade, L.~C.~T.~Guillen and J.~G.~Pereira,
       ``Gravitational energy momentum density in teleparallel gravity,''
       Phys. Rev. Lett. \textbf{84} (2000), 4533-4536
       
      
    
     
      \bibitem{ref28a}
     J.~Beltr\'an Jim\'enez, L.~Heisenberg, T.~S.~Koivisto and S.~Pekar,
     ``Cosmology in $f(Q)$ geometry,''
     Phys. Rev. D \textbf{101} (2020) no.10, 103507
     
      \bibitem{ref28b}     
     F.~G.~Alvarenga, M.~J.~S.~Houndjo, A.~V.~Monwanou and J.~B.~C.~Orou,
     ``Testing some $f(R,T)$ gravity models from energy conditions,''
     J. Mod. Phys. \textbf{4} (2013), 130-139
     
      \bibitem{ref28c}
      A.~K.~Mishra, U.~K.~Sharma, V.~C.~Dubey and A.~Pradhan,
      ``Traversable Wormholes in $f(R,T)$ Gravity,''
      Astrophys. Space Sci. \textbf{365} (2020) no.2, 34
     
    
     \bibitem{ref28d} 
     P.~Pavlovic and M.~Sossich,
     ``Wormholes in viable $f(R)$ modified theories of gravity and Weak Energy Condition,''
     Eur. Phys. J. C \textbf{75} (2015), 117
     
     \bibitem{ref29}  
      S.~V.~Sushkov and S.~M.~Kozyrev,
       ``Composite vacuum Brans-Dicke wormholes,''
       Phys. Rev. D \textbf{84} (2011), 124026
       
        
    
     
       
       \bibitem{ref31}
       A.~K.~Mishra, Shweta and U.~K.~Sharma,
       ``Non-exotic wormholes in 4-D Einstein-Gauss-Bonnet gravity,''
       [arXiv:2106.04369 [gr-qc]].
       
      
    

\end{thebibliography}
\end{document}